\documentclass[conference]{IEEEtran}
\IEEEoverridecommandlockouts
\usepackage{cite}
\usepackage{amsmath,amssymb,amsfonts}
\usepackage{algorithmic}
\usepackage{graphicx}
\usepackage{textcomp}
\usepackage{xcolor}
\def\BibTeX{{\rm B\kern-.05em{\sc i\kern-.025em b}\kern-.08em
		T\kern-.1667em\lower.7ex\hbox{E}\kern-.125emX}}
\begin{document}
	\title{Cross-layer Modeling and Design of Content Addressable Memories in Advanced Technology Nodes for Similarity Search }

	\author{Siri~Narla\thanks{This work was supported by COCOSYS, JUMP, an SRC program sponsored by DARPA.},
		~Piyush~Kumar\thanks{Siri Narla, Piyush Kumar, Mohammad Adnaan, and Azad Naeemi are with the School of Electrical and Computer Engineering, Georgia Institute of Technology, Atlanta, GA 30332 USA, email: pkumar315@gatech.edu.}, ~Mohammad~Adnaan, and~Azad~Naeemi}
	
	\maketitle
	
	\begin{abstract}
		In this paper we present a comprehensive design and benchmarking study of Content Addressable Memory (CAM) at the 7nm technology node in the context of similarity search applications. We design CAM cells based on SRAM, spin-orbit torque, and ferroelectric field effect transistor devices and from their layouts extract cell parasitics using state of the art EDA tools. These parasitics are used to develop SPICE netlists to model search operations. We use a CAM-based dataset search and a sequential recommendation system to highlight the application-level performance degradation due to interconnect parasitics. We propose and evaluate two solutions to mitigate interconnect effects.
	\end{abstract} 
	
	\begin{IEEEkeywords}
		Content Addressable Memory, Similarity Search, SOT-MRAM, ferroelectric, Interconnect, Parasitics, Recommendation Systems
	\end{IEEEkeywords}
	
	%
	\IEEEpeerreviewmaketitle
	
	\section{Introduction}
	%
	%
	%
	%
	\IEEEPARstart{I}{n} modern applications, datasizes can be large in both dimensions (D) and number of samples (n) that makes similarity search increasingly challenging \cite{tong2008lessons}. Thus, hardware solutions like Content Addressable Memories (CAMs), that can perform parallel in-memory search over an entire database, are being studied with great interest \cite{narla2022design, narla2022modeling,  yin2018ultra}. The speed-up in search they offer and their linear space requirements have made CAMs suitable for applications such as memory augmented neural networks \cite{ni2019ferroelectric}, hyper-dimensional computing \cite{HD_intro}, recommendation systems \cite{iMARS} ,and dataset searches. 
	
	\par
	The closest match in CAM is found by using Hamming distance (HDist) which is defined as the number of mismatching bits between two vectors. In most CAM designs, a matchline is precharged before every evaluation cycle. During evaluation, if there is a mismatch between the stored bit in a cell and the search bit, the cell discharges the matchline (ML). The discharge rate of ML increases with the number of mismatching bits in that row. This timing characteristic is used to represent the HDist between the query vector and the stored data. Despite its simplicity, there are challenges in using CAM for similarity search operations. Implementing a CAM cell with SRAM cells \cite{pagiamtzis2006content} requires 10 transistors which limits the memory capacity. SRAM also suffers from leakage which can become a major limiting factor for large datasets. To address these challenges, CAM cells based on emerging non-volatile memory (NVM) devices such as resistive memories, spin-orbit torque (SOT) devices, and ferroelectric field effect transistors (FeFETs) have been proposed \cite{li20131mb, narla2022design, yin2018ultra}. Each of these technology options come with their own set of advantages and limitations. FeFETs are voltage-controlled and can be quite energy efficient and compact. However, they generally suffer from poor endurance \cite{yurchuk2016charge} and currently require large write voltages (~4V) \cite{choe2021variability} and write times (>10ns) \cite{dunkel2017fefet}. SOT devices have a relatively low write delay and theoretically have unlimited endurance. However, they need large write energy due to their current based write scheme \cite{narla2022design}. Hence, the optimal technology choice is application-dependent.
	
	\begin{figure*}[h]
		\centering
		\includegraphics[width=\linewidth]{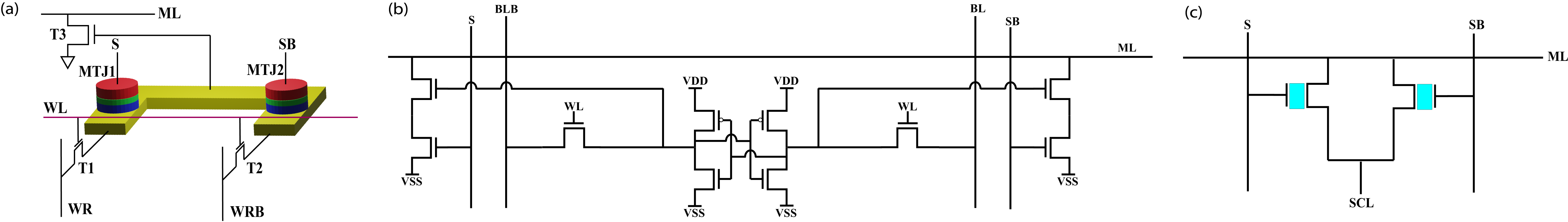}
		\caption{Cell schematics of (a) SOT-based CAM, (b) SRAM-based CAM and (c) FeFET-based CAM}
		\label{fig_cam_sch}
	\end{figure*}
	
	\begin{figure}[h]
		\centering
		\includegraphics[width=0.8\linewidth]{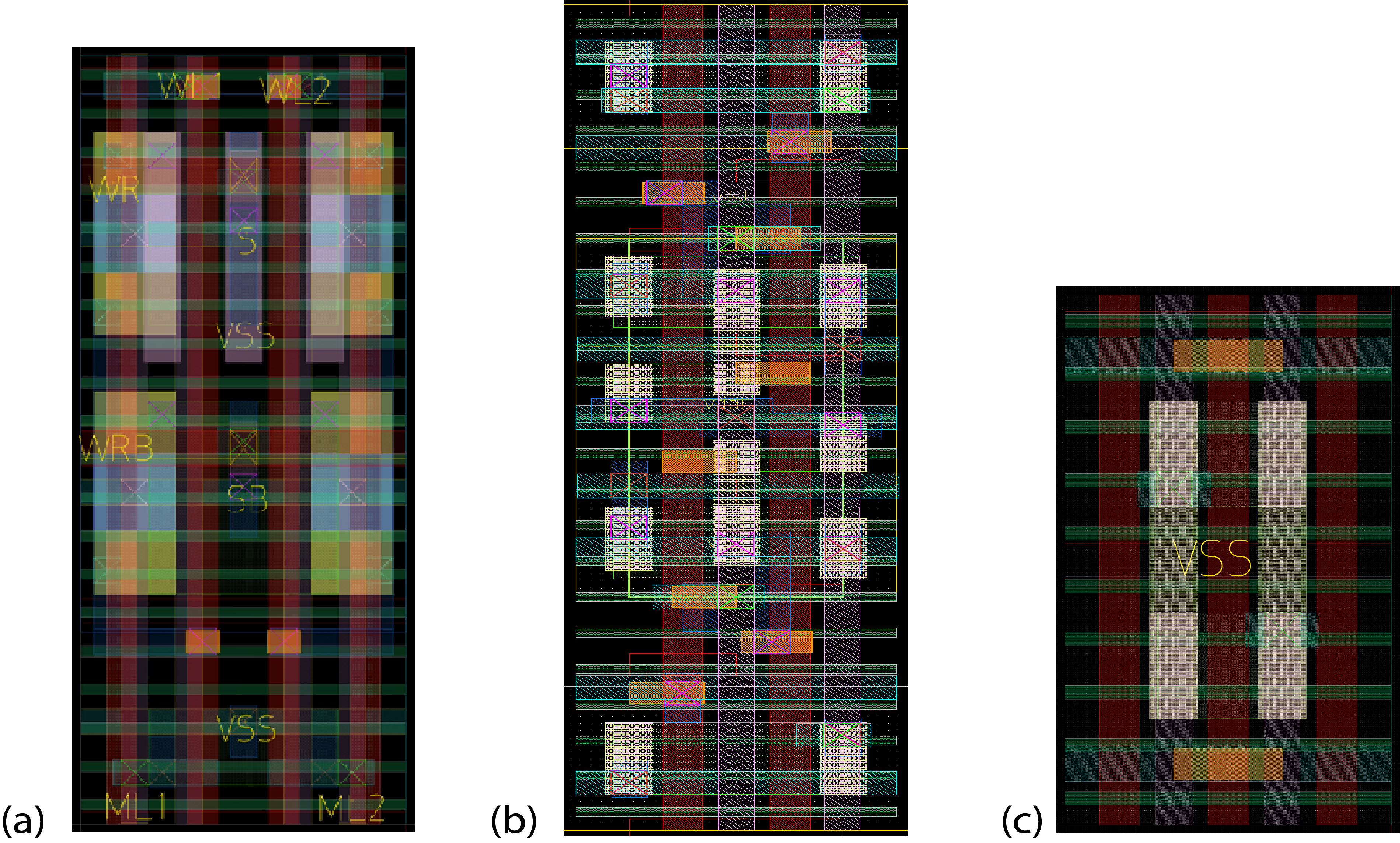}
		\caption{(a)SOT-CAM 2x1 cell layout. This layout shows 2 aligned cells that share WR, WRB and VSS contacts to reduce overall area. Cell size is 6Fx30F (b) SRAM based CAM cell layout. Cell size is 9Fx37.5F (c) FeFET-based CAM cell layout. Cell size is 9Fx15F. F is half M1 metal pitch and is 18nm.}
		\label{fig_cam_layout}
	\end{figure}
	
	\par
	NVM-based CAMs have been studied for a variety of applications \cite{ni2019ferroelectric, HD_intro, iMARS, imani2019searchd, narla2022modeling} that show major improvement in latency and energy compared to other conventional approaches. However, all these designs have been implemented at older technology nodes (45nm) where interconnect resistances were not as large as they are in more advanced nodes \cite{huang2023comprehensive}. Also, the area advantage offered by NVM-based CAMs over SRAMs shrink as technology advances \cite{sram_7nm}. To justify the wide-scale adoption of NVM-based CAMs, their potential performance at the advanced technology nodes must be rigorously analyzed based on their physical layouts and accurate interconnect models. With the rapid increase in wire and via resistances as technology advances, interconnect parasitics introduce timing variation that can drastically affect their search operations particularly for large CAM arrays. Match lines with the same Hamming distance can take different amount of time to discharge depending on their position in the array which may disrupt one to one mapping of ML discharge delay to Hamming distance.
	
	\par
	In this paper we present a comprehensive design and benchmarking study of CAM arrays based on CMOS and beyond-CMOS devices at the 7nm technology node in the context of similarity search applications. We design CAM cells based on SRAM, SOT, and FeFET devices and from their layouts extract their parasitics using state of the art EDA tools. We then develop SPICE netlists based on the extracted parameters and model their search operations accounting for several sources of variability. In addition, we propose techniques to mitigate the performance degradation due to interconnects and quantify their impact at the application-level for dataset searches and a recommendation model. 
	
	\par
	The paper is organized as follows. In Section 2 we briefly discuss the implementation of similarity search using SOT, SRAM, and FeFET-based CAMs. Section 3 discusses the challenges of implementing similarity search. In Section 4 we study the effects of interconnect parasitics on similarity search using various metrics. We also discuss potential solutions to work around this issue. In Section 5 we evaluate and benchmark the results for CAM-based search at the application-level for dataset search and recommendation systems. Finally, Section 6 concludes the paper.

	\section{Cell Design and Layout}
	To implement similarity search with CAMs we consider a 3 transistor, 2 SOT-driven MTJ-based CAM design \cite{narla2022design} (SOT-CAM) as shown in Fig \ref{fig_cam_sch} (a), a 10 transistor SRAM-based CAM design \cite{pagiamtzis2006content} (SRAM-CAM) as shown in Fig \ref{fig_cam_sch} (b) and a 2 FeFET-based CAM design \cite{yin2018ultra} (FeFET-CAM) as shown in Fig \ref{fig_cam_sch} (c). In all three cases, complementary data is stored in a CAM cell, matchlines are precharged to Vdd before evaluation and during search the searchlines, S and SB, are driven to Vs/0 depending on the search data.
	
	\par
	In the case of SOT-CAM, Vs gets divided between the two MTJ resistances and the resulting voltage, Vsot, drives the gate of the discharge transistor (T3). T3 switches on and discharges ML only in the case of a mismatch between the stored bit and the search bit. Likewise, in the SRAM-CAM, if the data stored in the SRAM mismatches  the search bit, then a path to ground is created and the ML is discharged. In an FeFET-CAM, Vs is chosen between the two possible threshold voltages of the FeFET and the precharged ML gets discharged if Vs is applied to the FeFET with the lower threshold voltage which happens in the case of a mismatch. We use an inverting sense amplifier (SA) to detect ML discharge.
	
	\begin{figure}
		\centering
		\includegraphics[width=0.8\linewidth]{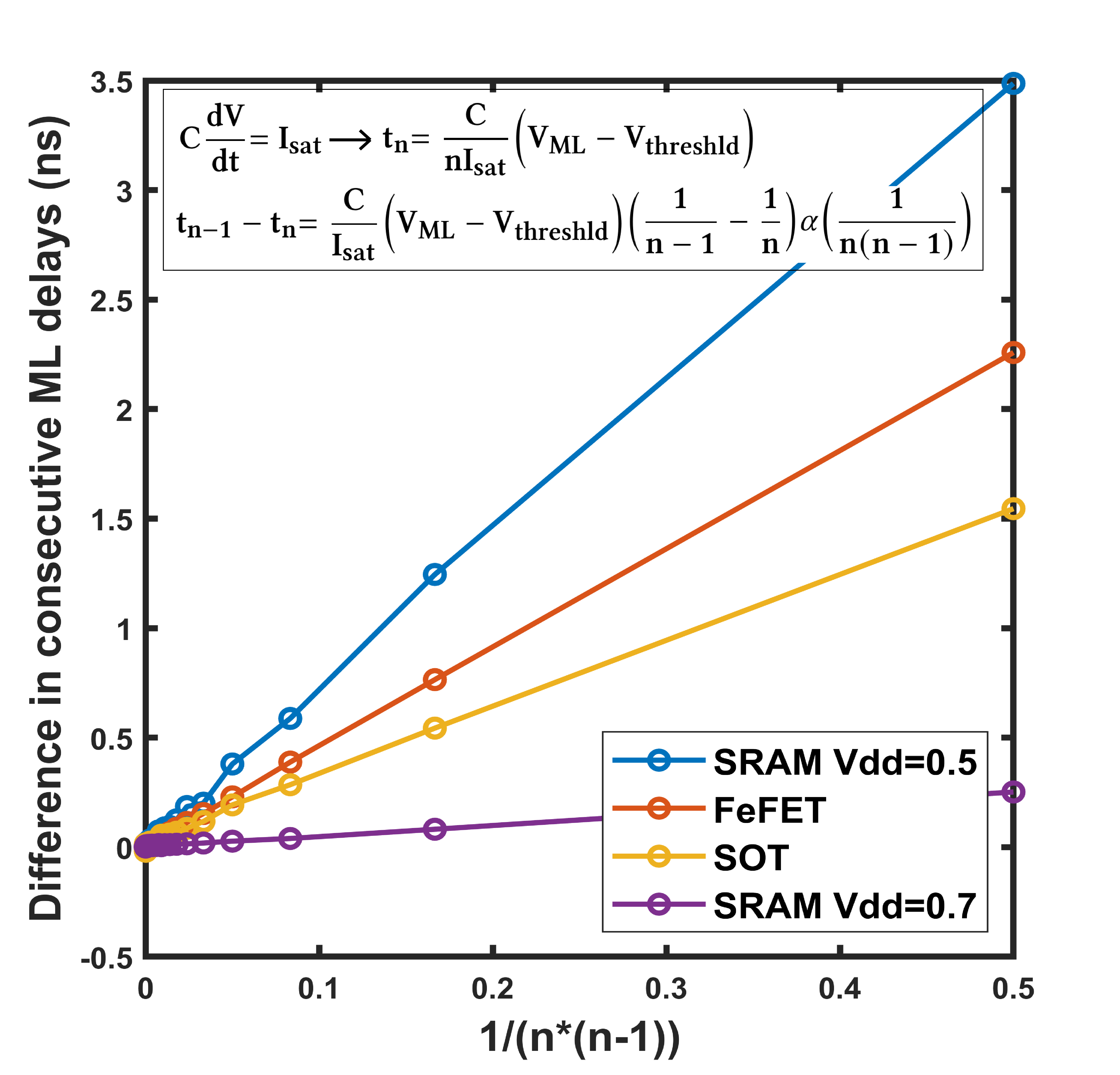}
		\caption{ The difference in ML discharge delay for HDist=n ( $\mathrm{t_{n}}$) and HDist=n-1 ( $\mathrm{t_{n-1}}$) vs 1/n(n-1). $\mathrm{V_{ML}}$ and $\mathrm{V_{threshold}}$ are matchline precharge voltage and voltage where the sense amplifier detects discharge. C is the ML capacitance and $\mathrm{I_{sat}}$ is the discharge current from a mismatching cell. }
		\label{fig_cam_ml_dis}
	\end{figure}
	
	\begin{figure*}[h]
		\centering
		\includegraphics[width=0.8\linewidth]{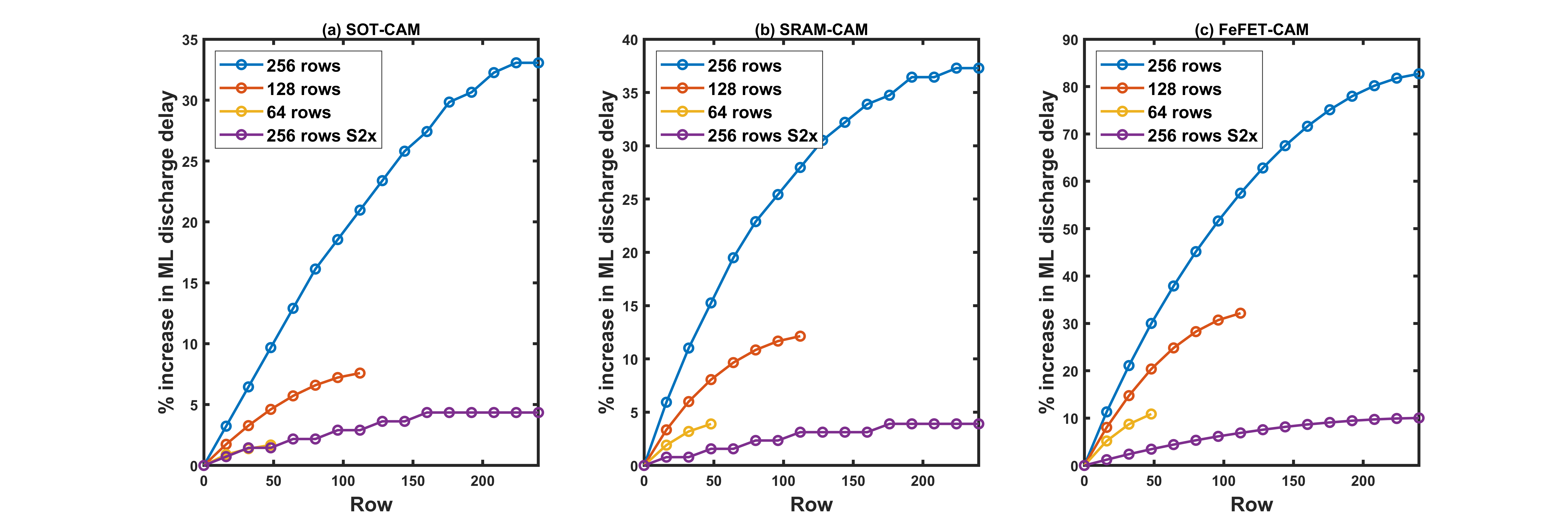}
		\caption{Percentage increase in delay to discharge MLs with HDist=40 from the row closest to the driver to the row farthest from the driver for (a) SOT-CAM array (b) SRAM-CAM array for Vdd=0.5V and (c) FeFET-CAM arrays with  128 columns and various number of rows. S2x is the case where the searchlines are shifted to M4 and made 2x wider.}
		\label{fig_rc_delay}
	\end{figure*} 	
	
	\par
	We extract the circuit parasitics based on our SOT, SRAM, and FeFET-CAM layouts shown in Fig \ref{fig_cam_layout} based on the ASAP7 PDK \cite{clark2016asap7}. The BEOL resistance values are obtained from TCAD simulations in \cite{huang2023comprehensive} and are used to generate the nxtgrd database containing the resistance and capacitance information for various layers. Then, the nxtgrd database is used in Synopsys’ StarRC for extracting the parasitic resistance and capacitance from the layouts. These parasitics are used in SPICE simulations to extract ML discharge delays for various array sizes, Hamming distance values and row positions. For SOT-CAMs we use MTJs with 45nm diameter and 2nm oxide thickness. The tunnel magnetoresistance (TMR), resistance-area product and half bias values are obtained from the experimental results in \cite{yuasa2004giant}. The TMR degradation due to bias voltage is also incorporated as shown in \cite{narla2022design}.  For the 7nm FeFET-CAM, we use an FeFET with a memory window of 0.46V which has been reported in \cite{choe2021variability}. The capacitance of the ferroelectric layer is calculated considering the ferroelectric layer thickness of 5nm and a dielectric coefficient of 35 \cite{hsu2020theoretical}.

	\section{Search Performance Modeling}
	The accuracy and resolution of similarity search using CAMs are limited by several factors. As  the number of mismatching cells along a row increases, the ML discharge rate increases. However, the ML discharge delay is not linearly proportional to the number of mismatches in a row. In our CAM designs, the discharging transistor is mostly in the saturation mode before the ML discharge is detected by the SA. Therefore, ignoring the short-channel effects, the difference in ML discharge delays for HDist=n, $\mathrm{t_{n}}$ and HDist=(n-1), $\mathrm{t_{n-1}}$ is approximately proportional to 1/(n(n-1)) (Fig \ref{fig_cam_ml_dis}). This approximate relationship is validated by SPICE simulations in Fig \ref{fig_cam_ml_dis} and can be used to qualitatively explain some of the results in later sections. As HDist increases the difference in ML discharge delays between consecutive HDist values shrinks. Likewise, if the ML capacitance is smaller or the discharge current per mismatching cell ($\mathrm{I_{sat}}$) is larger, the delay separation drops. Hence, to improve the resolution of SRAM-CAM we lower search voltages and reduce hold voltages to 0.5V to lower $\mathrm{I_{sat}}$ compared to the standard case of Vdd=0.7V as seen in Fig \ref{fig_cam_ml_dis}. Also, the lower holding voltage reduces energy dissipation for SRAM-CAMs.
	
	\par
	Interconnect parasitics can also degrade the accuracy of similarity search due to IR drop and RC delay especially if the array size is large. Fig \ref{fig_rc_delay} shows the percentage increase in delay from the row closest to the driver to the row farthest from the driver for a fixed HDist of 40. In the case of SRAM, and FeFET-CAMs (Fig \ref{fig_rc_delay} b and c), the variability is mainly the result of the RC delay to charge S/SB lines. In the case of SOT-CAMs (Fig \ref{fig_rc_delay} (a)) a DC current passes through the MTJs that  results in an IR drop on the S/SB lines which lowers the gate voltage of the discharge transistor as the distance to the driver increases.
	\par
	To reduce the IR drop and RC delay in large CAM arrays, the searchlines (S and SB) can be moved from M2 to M4, where they can be made 2x wider. It can be seen in Fig \ref{fig_rc_delay} (S2x) that this can provide 7.6x, 9.5x and 8.2x reductions in delay variability for SOT, SRAM, and FeFET arrays with 256 rows, respectively. However, this comes at the cost of a higher S/SB line capacitance and more complex CAM cells as will be discussed later. 
	\par
	Inconsistency in ML discharge delays combined with the shrinking difference in discharge delays for larger consecutive HDists, introduce a major challenge in accurately mapping Hamming distances to ML discharge delays.

	\begin{figure*}[h]
		\centering
		\includegraphics[width=0.8\linewidth]{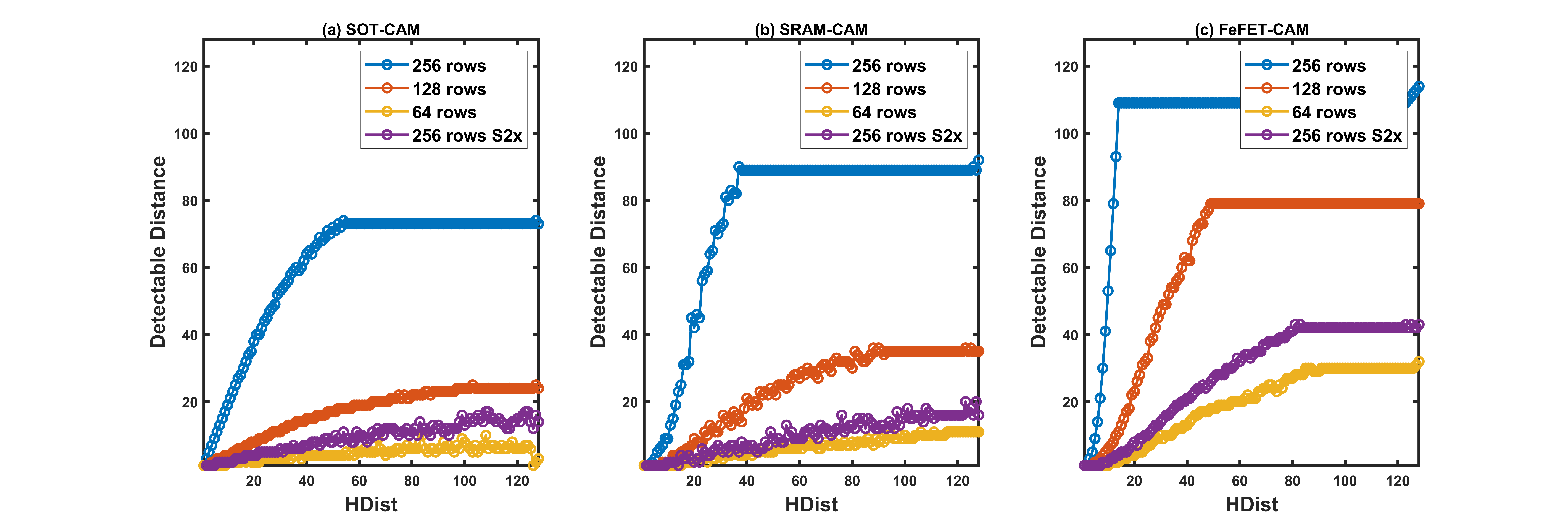}
		\caption{Minimum detectable distance for (a) SOT-CAM, (b) SRAM-CAM, and (c) FeFET-CAM arrays.}
		\label{fig_detectable}
	\end{figure*}
	
	\section{Search Metrics}
	
	Here we define and quantify a few important metrics for search operations to analyze and mitigate the limitations pointed out in Section 3.
	
	\subsection{Minimum Detectable Distance (Resolution)}
	\par
	Search resolution is the ability to clearly distinguish between rows with Hamming distance values that are close to each other. This can be measured in terms of the minimum detectable distance (MDD). Two HDist values can be distinguished if their corresponding delays are distinct irrespective of their positions in the array.  MDD is plotted in Fig \ref{fig_detectable} versus HDist for various array sizes and CAM technology options. As the array size and HDist increase, MDD increases due to more pronounced interconnect effects (shown in Fig \ref{fig_rc_delay}) and the shrinking difference between the delays of consecutive hamming distances when wire parasitics are ignored (shown in Fig \ref{fig_cam_ml_dis}). 
	
	\par
	SOT-CAMs offer a better resolution than SRAM-CAMs at larger HDist because ML delay variation across the array is larger for SRAM-CAMs as seen in Fig \ref{fig_rc_delay} (b). In the case of the FeFET-CAM, the FeFET capacitance contributes significantly to the RC delay . Also, the cell width of the FeFET-CAM is smaller than those of the SOT and SRAM-CAMs which results in a smaller ML capacitance; hence, a smaller delay separation (Fig \ref{fig_cam_ml_dis}). These two factors contribute to the drop in resolution seen in Fig \ref{fig_detectable} (c)). 
	
	\subsection{Precision, Recall Rate and F-scores}
	The metrics precision, recall rate and f-scores are defined in Table \ref{tab_def} in the context of fixed-radius near neighbor search which is the search of all items with an HDist smaller than a given value (HDist limit). To perform this kind of search we can implement delay thresholding by using a latch after each sense amplifier and control the timing of the edge of the clock (Clk). 
	
	\begin{table}
		\caption{Precision, Recall Rate and F-score definition.}
		\label{tab_def}
		\begin{tabular}{cl}
			Precision & Percentage of relevant items in the\\&  retrieved items \\ 
			Recall Rate   & Percentage of relevant items that were retrieved       \\ 
			F-score   & Harmonic mean of precision and recall                  \\ 
		\end{tabular}
	\end{table}
	\begin{table}
		\caption{Fixed-radius Near Neighbor Search Results for SOT-CAM.}
		\label{tab_nns_sot}
			\begin{tabular}{cllllll}
				&  &  &  &  & S2x & Clk match  \\ 
				Row size & Ideal & 64 & 128 & 256 & 256 & 256  \\ 
				Recall Rate & 100 & 100 & 100 & 100 & 100& 100  \\ 
				Precision & 100 & 98 & 83.8 & 49.1 & 91.76 & 86 \\ 
				F-score & 1 & 0.99 & 0.91 & 0.66 & 0.96& 0.92\\ 
				EPS (pJ) & & 1.88 & 3.55  & 7.16 & 8.22& 7.34            \\
				Delay (ns)  & & 1.2   & 1.2    & 1.6  & 1.6& 1.6     \\ 
			\end{tabular}
		\end{table}
		\begin{table}
			\caption{Fixed-radius Near Neighbor Search Results for SRAM-CAM.}
			\label{tab_nns_sram}
			\begin{tabular}{cllllll}
				&  &  &  &  & S2x & Clk match  \\ 
				Row size & Ideal & 64 & 128 & 256 & 256 & 256  \\ 
				Recall Rate & 100 & 100 & 100 & 100 & 100& 93.4  \\ 
				Precision & 100 & 100 & 83 & 45.5 & 96.1 & 92.6\\ 
				F-score & 1 & 1 & 0.91 & 0.62& 0.98& 0.93\\ 
				EPS (pJ) & & 0.8 & 1.7  & 3.33 & 3.44&      3.65    \\
				Delay (ns)  & & 0.8  & 1.2    & 1.6  & 1.6  & 1 .6  \\ 
			\end{tabular}
		\end{table}
		\begin{table}
			\caption{Fixed-radius Near Neighbor Search Results for FeFET-CAM.}
			\label{tab_nns_fefet}
			\begin{tabular}{cllllll}
				&  &  &  &  & S2x & Clk match  \\ 
				Row size & Ideal & 64 & 128 & 256 & 256 & 256  \\ 
				Recall Rate & 100 & 100 & 100 & 100  & 100 & 100 \\ 
				Precision & 100 & 93.13 & 65.22 & 24.4  & 83.5& 70.7\\ 
				F-score & 1 & 0.96 & 0.79 & 0.39 & 0.91  & 0.83\\ 
				EPS (pJ) & & 0.78 & 1.66 & 3.15   & 3.95   &   3.28   \\
				Delay (ns)  &  &0.4   & 0.8    & 1.2    & 1.2   &  1.2 \\ 
			\end{tabular}
		\end{table}
		
		\par
		Tables \ref{tab_nns_sot}, \ref{tab_nns_sram} and \ref{tab_nns_fefet} show the results for CAMs with 10000 items with 128 bit binary data initialized with random Hamming distances for SOT, SRAM, and FeFET-CAMs, respectively for a Hdist limit of 20 along with the energy (EPS) and delay values per search operation. For all technology options, the quality of search drops as the number of rows increases due to interconnect effects.  Moving search lines to M4 (S2x) can significantly improve the metrics at the cost of 14.8\%, 3.3\% and 25.4\% increase in energy per search operation for SOT, SRAM, and FeFET-CAMs, respectively due to higher S/SB line capacitance. Alternatively, we also look at the case where the RC delay on the Clk lines is matched with the RC delay on the search lines (Clk match) by increasing the number of fins of the transistors in the latch that are driven by the Clk from 2 to 6. This method significantly improves precision (Table \ref{tab_nns_sot} and \ref{tab_nns_fefet}). The improvement from this method depends on how well Clk RC delay matches the RC delay for a given HDist limit and can sometimes cause a drop in recall rate as seen in Table \ref{tab_nns_sram}.

		\begin{table}
			\caption{With and without variation, fixed-radius near neighbor search results over 10000 128 bit items stored in SOT-CAM arrays of size 128x128.}
			\label{tab_nns_var}
			\begin{tabular}{p{45pt}p{38pt}p{38pt}p{38pt}p{38pt}}
					& Without Variation & With Variation & Without Variation & With Variation  \\
					HDist limit & 20 & 20 & 5 & 5 \\
					Recall  Rate  & 100   & 98.15   & 100 &   96.88      \\
					Precision & 83.8   & 84.84 & 93.87 & 86.55    \\
					F-score   & 0.91     & 0.91 & 0.97  & 0.91\\
				\end{tabular}
			\end{table}
			
			\par
			To understand the effect of variation, we pick SOT-CAM and run 100 Monte Carlo simulations for each Hamming distance value for various row positions in the CAM array assuming a Gaussian distribution with 3$\sigma$ MTJ resistance variation of 15\% \cite{everspin,sensing_rev} and 3$\sigma$ threshold voltage variation of 42mV \cite{giles2015high}. Our results in Table \ref{tab_nns_var} for a 128x128 array show that the results are not too sensitive to variability especially when larger HDist limit values are considered.
			\par
			SOT-CAMs (Table \ref{tab_nns_sot}) dissipate more energy because unlike the FeFET (Table \ref{tab_nns_fefet}) and SRAM-CAMs (Table \ref{tab_nns_sram}), a DC current flows through the SOT-CAM circuit during search. As the number of rows increase, the RC delay from the driver increases, which increases the search delay for all the designs. While S2x improves search performance, it comes at the cost of more energy to charge the searchlines due to an increase in capacitance. Using Clk matching increases search energy slightly as energy to charge the Clk line increases.
			
			\section{Application-level Evaluation and Results}
			To understand the effect of interconnect parasitics at the application-level, we look at a dataset search and a sequential recommendation system where the similarity search is implemented using a CAM. 
			\par
			We use the Kaggle Housing dataset \cite{house-dataset} to implement fixed-radius near neighbor search over the dataset. The dataset has 80 features and 1460 items. For the purpose of our experiment we use 16 features with less than 8 unique values and one hot encode this categorical data using 8 bits per feature. Table \ref{tab_housing_sot}, \ref{tab_housing_sram} and \ref{tab_housing_fefet} show that increasing the number of rows comes at a significant cost of 2.5x, 1.75x, 2.66x  drop in precision for SOT, SRAM ,and FeFET-CAM arrays respectively. Using the S2x (Clk match) solution to mitigate interconnect effects improves f-scores for SOT, SRAM, and FeFET-CAMs with 256 rows by 1.88x (1.36x), 1.43x (1.36x) and 1.96x (1.65x) respectively. Results are averaged over 100 random queries with HDist limit=8. 
			
			\begin{table}
				\caption{Search Results for Housing dataset with HDist limit=8 using SOT-CAMs. }
				\label{tab_housing_sot}
				\begin{tabular}{cllllll}
					&  &  &  &  & S2x & Clk match  \\ 
					Row size & Ideal & 64 & 128 & 256 & 256 & 256  \\ 
					Recall Rate & 99.7 & 99.6 & 99.8 & 99.8 & 99.7 & 99.75  \\ 
					Precision & 100 & 100 & 74 & 40 &  100 & 58  \\ 
					F-score & 1 & 1 & 0.84 & 0.53 & 1 & 0.72\\ 
				\end{tabular}
			\end{table}
			\begin{table}
				\caption{Search Results for Housing dataset with HDist limit=8  using SRAM-CAMs. }
				\label{tab_housing_sram}
				\begin{tabular}{cllllll}
					&  &  &  &  & S2x & Clk match  \\ 
					Row size & Ideal & 64 & 128 & 256 & 256 & 256  \\ 
					Recall Rate & 99.7 & 99.8 & 99.7 & 99.8  & 99.7 & 90\\ 
					Precision & 100 & 100 & 100 & 57 & 100  & 100\\ 
					F-score & 1 & 1 & 1& 0.7 &1& 0.95\\ 
				\end{tabular}
			\end{table}
			\begin{table}
				\caption{Search Results for Housing dataset with HDist limit=8  using FeFET-CAMs.}
				\label{tab_housing_fefet}
				\begin{tabular}{cllllll}
					&  &  &  &  & S2x & Clk match  \\ 
					Row size & Ideal & 64 & 128 & 256 & 256 & 256  \\ 
					Recall Rate & 99.8 & 99.8 & 99.78 & 99.68  & 99.8 & 99.8\\ 
					Precision & 100 & 100 & 75.62 & 37.54 & 100  & 73.4\\ 
					F-score & 1 & 1 & 0.85& 0.51 &1& 0.84\\ 
				\end{tabular}
			\end{table}
			
			\par
			For the recommendation system (RS) we use the sequential recommendation model from \cite{sasrec}. In this model a self-attention block is used to predict the embedding of the next item that should be recommended to a user. In \cite{sasrec}, they use dot product ranking (DPR) to find the top-k items that are closest to the predicted item embedding. Using CAMs for ranking can be very costly as it would require ML discharge delay ranking to rank the top-k items which needs significant amount of peripherals. We use a concept similar to what was suggested in \cite{iMARS} where CAMs are used for candidate generation with fixed-radius near neighbor search and these candidates are passed to the ranking stage to rank the top-k items. In this way one can significantly reduce the number of items that need to go through DPR which is computationally expensive. We use the MovieLens 1M dataset from \cite{movielens} to train and test our model. Item embeddings are stored in the CAM using LSH encoding \cite{ni2019ferroelectric} with 128 bits. The attention model was trained for 100 epochs. During inference we used a test case with 1000 randomly selected negative items and 1 ground truth next item using the strategy from \cite{koren2008factorization}. HR@10 counts the fraction of times that the ground-truth next item is among the top 10 items after ranking for all valid test users. Our results show that with the use of CAMs, the final HR@10 (Table \ref{tab_sasrec_sot}) achieved are the same as those using DPR for the entire test set while requiring fewer dot product operations. 
			
			\begin{table}
				\caption{Results of Sequential RS with SOT-CAM Search. }
				\label{tab_sasrec_sot}
				\begin{tabular}{cllllll}
					&  &  &  &  & S2x & Clk match \\ 
					Row size & Ideal & 64 & 128 & 256 & 256 & 256  \\ 
					Mean pool size & 141.7 & 218.2 & 480.9 & 860.5 & 302.7 & 164.6\\ 
					CAM HR@10 & 0.32 & 0.32 & 0.32 & 0.32 & 0.32 & 0.31  \\ 
					Baseline HR@10 & 0.32 & 0.32 & 0.32 & 0.32 & 0.32 & 0.32 \\ 
					DPR reduction & 7.1x & 4.6x & 2.1x & 1.2x & 3.3x & 6.1x\\
				\end{tabular}
			\end{table}
			\begin{table}
				\caption{Results of Sequential RS with SRAM-CAM Search. }
				\label{tab_sasrec_sram}
				\begin{tabular}{cllllll}
					&  &  &  &  & S2x & Clk match \\ 
					Row size & Ideal & 64 & 128 & 256 & 256 & 256  \\ 
					Mean pool size & 141.8 & 219.5 & 661.7 & 909.9& 323.8& 368\\ 
					CAM HR@10 & 0.32 & 0.32 & 0.32 & 0.32 & 0.32 & 0.3 \\ 
					Baseline HR@10 & 0.32 & 0.32 & 0.32 & 0.32 & 0.32 & 0.32  \\ 
					DPR reduction & 7.1x & 4.6x & 1.5x & 1.1x& 3.1x & 2.7x \\
				\end{tabular}
			\end{table}
			\begin{table}
				\caption{Results of Sequential RS with FeFET-CAM Search.  }
				\label{tab_sasrec_fefet}
				\begin{tabular}{cllllll}
					&  &  &  &  & S2x & Clk match \\ 
					Row size & Ideal & 64 & 128 & 256 & 256 & 256  \\ 
					Mean pool size & 141.6 & 484.7 & 852.5 & 941.6& 701 & 737.5\\ 
					CAM HR@10 & 0.32 & 0.32 & 0.32 & 0.32 & 0.32 & 0.32 \\ 
					Baseline HR@10 & 0.32 & 0.32 & 0.32 & 0.32 & 0.32 &0.32 \\ 
					DPR reduction & 7.1x & 2.1x & 1.2x & 1.1x & 1.43x  & 1.36x\\
				\end{tabular}
			\end{table}
			
			\par
			Tables \ref{tab_sasrec_sot}, \ref{tab_sasrec_sram}, and \ref{tab_sasrec_fefet} show the results of using SOT, SRAM, and FeFET-CAMs for fixed-radius near neighbor search for candidate generation for various numbers of rows. In the ideal case we see on average almost 7.1x reduction in the number of items that need to undergo DPR (mean pool size). However, when interconnect parasitics are properly accounted for, increasing the number of rows in the CAM array significantly increases the number of candidates generated. Larger pool sizes increase the energy costs since more items have to go through DPR. By implementing our solutions S2x (Clk match), the size of the pool for a 256 row SOT, SRAM, and FeFET-CAM array can be brought down by 2.84x (5.2x), 2.8x (2.47x) and 1.34x (1.28x), respectively without significantly compromising the hit rate. 
			\par
			A dot product operation between two vectors of size R(1xm) requires m multiplications and (m-1) additions. A single mode processor can typically perform one addition and one multiplication per cycle \cite{tan2023low}. For n items, the number of cycles needed to compute the dot product increases by n times. Ranking to find top-k has a time complexity of O(nlogk). Overall, if CAMs can reduce the number of items that need to undergo DPR by 4.6x (Table \ref{tab_sasrec_sot} for 64 rows) with parallel search through the CAM in 3 cycles (Table \ref{tab_nns_sot}), then we can gain a 4.6x improvement in speed for the RS. Using multiple processors for parallel processing can reduce the number of cycles required to perform DPR; however, it will come at the cost of an area overhead \cite{tan2023low}. 
			
			\section{Conclusion}
			A cross-layer design and benchmarking framework for CMOS and beyond-CMOS CAMs at the 7nm technology node is presented highlighting the major impact of interconnects on the quality of search in CAMs at the advanced technology nodes. We show that the limits imposed by interconnects can get partially mitigated by either using upper metal levels for search lines or balancing the RC delay of the clock lines with those of the search lines. For SRAM-CAM lowering voltages to 0.5V can significantly improve search performance. We also show that SOT and FeFET-CAMs provide 3.75x and 2.5x improvement in terms of desnsity with respect to SRAM, at comparable speed. While both SOT and FeFET-CAMs eliminate the standby leakage power, the SOT-CAM dissipates 2x dynamic energy per search operation compared to FeFET and SRAM-CAMs. At the application-level, we show that as the array size  increases from 64x128 to 256x128, f-scores for a CAM-based dataset search and reduction in dot product operations for a sequential recommendation system drop by 1.88x and 3.8x, respectively due to interconnect parasitics and the proposed solutions can improve these results by up to 1.88x and 2.75x, respectively.

			\bibliographystyle{IEEEtranU}
			\bibliography{sample-base}

		\end{document}